\documentclass[12pt]{article}
\usepackage{axodraw}
\newcommand{\al }{\alpha}
\newcommand{\gamu }{\gamma^\mu}
\newcommand{\Gamu }{\Gamma^\mu}
\newcommand{\ganu }{\gamma^\nu}
\newcommand{\gaal }{\gamma^\al}
\newcommand{\garo }{\gamma^\rho}
\newcommand{\gasi }{\gamma^\sigma}
\newcommand{\gafi }{\gamma_5}
\newcommand{\tr }[1]{\mathrm{Tr}(#1)}
\newcommand{\ac }[1]{\mathrm{AC}(#1)}
\newcommand{\trfi }{\tr{\gamu \ganu \garo \gasi \gafi}}

\newcommand{\be}{\begin{equation}}
\newcommand{\ee}{\end{equation}}
\newcommand{\ba}{\begin{eqnarray}}
\newcommand{\ea}{\end{eqnarray}}
\newcommand{\bea}{\begin{eqnarray*}}
\newcommand{\bet}{\begin{center} \begin{tabular}}
\newcommand{\eea}{\end{eqnarray*}}
\newcommand{\bary}{\begin{array}}
\newcommand{\eary}{\end{array}}
\newcommand{\ent}{\end{tabular} \end{center}}
\newcommand{\bb}{\begin{thebibliography}}
\newcommand{\eb}{\end{thebibliography}}

\newcommand{\ra}{\rightarrow}

\newcommand{\bit}{\begin{itemize}}
\newcommand{\eit}{\end{itemize}}
\newcommand{\eps}{\epsilon}
\newcommand{\veps}{\varepsilon}
\newcommand{\sla}[1]{#1 \!\!\!/}

\newcommand{\crn}{\nonumber \\}

\newcommand{\ha}{\frac{1}{2}}

\newcommand{\WTis}{Ward-Takahashi identities }
\newcommand{\WTisp}{Ward-Takahashi identities.}
\newcommand{\WTisc}{Ward-Takahashi identities,}

\newcommand{\npb}{{\em Nucl.\ Phys.\ }{\bf B}}
\newcommand{\np}{{\em Nucl.\ Phys.\ }{\bf B}}
\newcommand{\pl}{{\em Phys.\ Lett.\ }}
\newcommand{\prd}{{\em Phys.\ Rev.\ }{\bf D}}
\newcommand{\prl}{{\em Phys.\ Rev.\ Lett.\ } }
\newcommand{\pr}{{\em Phys.\ Rev.\ }}
\newcommand{\zp}{{\em Z.\ Phys.\ }{\bf C}}
\newfont{\grfett}{cmmib10 scaled 1100}
\newfont{\liste}{pzdr scaled 1100}
\begin{document}

\parindent 0mm
\parskip 2mm
\renewcommand{\arraystretch}{1.4}
\renewcommand{\thefootnote}{\fnsymbol{footnote}}
\thispagestyle{empty}
\hfill {\sc DESY 00-075} \qquad { } \par
%\hfill { hep-ph/0005yyy} \qquad { } \par
\hfill May 2000 \qquad { }

\vspace*{10mm}

\begin{center}

{\Large {\bf Facts of life with $\gafi$}\vglue 10mm }

\vspace{8mm}

{\large
 F. Jegerlehner,

\vspace{1cm}
Deutsches Elektronen-Synchrotron DESY \\
Platanenallee 6, D--15738 Zeuthen, Germany\\
}
\end{center}

\textwidth 120mm
\begin{abstract}
The increasing precision of many experiments in elementary particle
physics leads to continuing interest in perturbative higher order
calculations in the electroweak Standard Model or extensions of it.
Such calculations are of increasing complexity because more loops
and/or more legs are considered. Correspondingly efficient
computational methods are mandatory for many calculations. One problem
which affects the feasibility of higher order calculations is the
problem with $\gafi$ in dimensional regularization.  Since the subject
thirty years after its invention is still controversial I advocate
here some ideas which seem not to be common knowledge but might shed
some new light on the problem. I present arguments in favor of
utilizing an anticommuting $\gafi$ and a simple 4--dimensional treatment
of the hard anomalies.
\end{abstract}

\\
\textwidth 170 mm
\newpage

%\setcounter{page}1

%\renewcommand{\thefootnote}{\arabic{footnote}}
%\setcounter{footnote{0}}
%\hfill {\em Ne vous compliqu\'e pas la vie} \qquad { } \par
%\hfill { R. Guignard } \qquad { } \par
%\hfill { (my former french teacher)} \qquad { }
\section{Introduction}
The electroweak Standard Model (SM)~\cite{SM} has been extremely
successful in the interpretation of LEP/SLC data and higher order
effects typically amount to 10 $\sigma$ deviations if not taken into
account~\cite{Sirlin99}. These precise predictions are only possible
due to the renormalizability~\cite{tHooft71} of the SM and the by now
very precise knowledge of the relevant input parameters. Last but not
least the relevant coupling constants are small enough such that
perturbation theory mostly works very well.

The formal proofs of renormalizability of the SM~\cite{tHVR} often
relied on the assumption that a gauge invariant regularization
exists. The question whether such a regularization exists is
non--trivial because of the chiral structure of the fermions
involved. At present the only regularization, which makes elaborate
computations of radiative corrections feasible, is the dimensional
regularization (DR) scheme~\cite{tHVD,BGDR} which is well-defined for
field theories with vectorial gauge symmetries only. However, in
theories exhibiting chiral fermions, like the electroweak SM, problems
with the continuation of the Dirac matrix $\gafi$ to dimensions $D\neq
4 $ remain open within this context and several modifications of the 't
Hooft--Veltman DR have been
proposed~\cite{BGL,Akye,BM77,Siegel79,ChFH79,NaWe79,GB80,AT81,BDGKSch91}.
It
turns out that starting from the standard SM-Lagrangian and using a
$\gafi$, which does not anticommute with the other Dirac matrices
$\gamu $, leads to ``spurious anomalies'' which violate chiral
symmetry and hence gauge invariance. These anomalies would spoil
renormalizability if we would not get rid of them by imposing ``by
hand'' the relevant Ward-Takahashi (WT)~\cite{WT} and Slavnov-Taylor
(ST)~\cite{ST,BRS} identities order by order in perturbation
theory~\cite{GB80,KNSch90,BuWe90,La93,FLOR95}. At first sight this
might not look to be a serious problem, however, violating the
symmetries of the SM makes practical calculations much more difficult
and tedious than they are anyway.

The problems of course are related to the existence of the
Adler--Bell--Jackiw (ABJ) anomaly~\cite{ABJ}, which must cancel in the SM in
order not to spoil its renormalizability~\cite{SManomaly}.

Surprisingly, the prescriptions proposed and/or used by many authors
continue to be
controversial~\cite{BM77,ChFH79,GB80,BDGKSch91,BuWe90,GD79,ET82,Ovrut83,AZ87,cyclic,MP99,Wein00,Frei00},
and hence it seems to be necessary to reconsider the problem once
again. We shall emphasize, in particular, the advantage of working with
chiral fields. The consequences of working as closely as possible with
chiral fields, it seems to me, has not been stressed sufficiently in
the literature so far.

As a matter of principle it is important to mention two other
approaches which both work in $D=4$ dimensions. i) In quantum field
theories on the lattice a recent breakthrough was the discovery of
exact chiral invariance on the lattice~\cite{MLCHI} which circumvents
the Nielsen--Ninomiya no--go theorem~\cite{NNnogo}. A well defined
regularization which preserves simultaneously chiral--and
gauge--symmetries is thus known and could be applied to the SM. ii)
The algebraic renormalization of the electroweak SM to all
orders~\cite{EKraus} within the Bogoliubov-Parasiuk-Hepp-Zimmermann
(BPHZ) framework is a mathematically well defined scheme, which is
much more involved because it breaks the symmetries at intermediate
stages and hence leads to much longer expressions which are extremely
tedious to handle in practice. In cases of doubt this is the only
known scheme which is free of ambiguities and works directly in
4--dimensional continuum field theory.

For perturbative calculations in the continuum we have to stick as
much as possible to the more practical route of dimensional
regularization. In the following tensor quantities in $D=4$ dimensions
are supposed to be defined by interpolation of $D=2n$ ($n\geq 2$,
integer ) dimensions to dimensions below $D=4$. It is well
known that the $\gamma$--algebra, the so called ``naive dimensional
regularization'' (NDR)~\footnote{I read it as ``normal dimensional
regularization''}
\be
\label{NDR}
\{\gamu,\ganu\}=2g_{\mu\nu} \cdot \mathbf{1}\;,\;\;g^\mu_{~~\mu}=D\;,\;\;
\ac{\mu} \equiv \{\gamu,\gafi\}=0
\ee
for dimensions of space--time $D=4-2\eps$, $\eps \neq 0$ is
inconsistent with
\be
\label{trcond}
\trfi \neq 0\;.
\ee
The latter condition is often considered to be necessary, however, for
an acceptable regularization since at $D=4$ we must find
\be
\label{epsten}
\trfi = 4 \mathrm{i} \veps^{\mu \nu \rho \sigma} \;.
\ee
Generally, for $\gafi$ odd traces one obtains trace conditions from
the cyclic property of traces. They are not fulfilled automatically,
as we shall see, and hence the algebra is ill-defined in general.
Considering $\tr{\prod\limits_{j=0}^{4} \gamma^{\mu_{j}}
\gafi \gaal }$ cyclicity requires
\be
\label{trgene}
\tr{\prod\limits_{j=0}^{4} \gamma^{\mu_{j}} \ac{\al}}-
2\:\sum\limits_{i=0}^{4} (-1)^i g^{\al \mu_{i}}
\tr{\prod\limits_{j=0,\: j\neq i}^{4} \gamma^{\mu_{j}} \gafi}=0\;.
\ee
Contraction with the metric tensor $g_{\al \mu_0}$ yields
\be
2\:(g^\al_{~~\al}-4)\:\tr{\prod\limits_{j=1}^{4} \gamma^{\mu_{j}}
\gafi}+\tr{\prod\limits_{j=1}^{4} \gamma^{\mu_{j}} \ac{\gamma}}=0
\ee
with $\ac{\gamma}\equiv \gamma_\al \ac{\al}$. Thus
$g^\al_{~~\al}=D\neq 4$ together with (\ref{trcond}) implies
$\ac{\mu}\neq 0$. However, non-anti-commutativity of $\gafi$ is in
conflict with the chiral structure and hence with gauge invariance of
the SM, in general. It is the purpose of this note to study the
possibility of restoring gauge invariance by employing chiral fields
systematically.

\section{Formally gauge invariant Feynman rules}

Obviously only terms involving $\gamu$ in the standard SM Lagrangian
can be affected by a non--anticommuting $\gafi$. As an example we
consider the leptonic part, given by
\ba
{\cal L}_\ell &=&
 \bar{\ell}_{\mathrm{R}} i \gamu \:(\partial_\mu+ig'B_\mu)\ell_{\mathrm{R}}
+\bar{\nu}_{\ell\mathrm{R}} i \gamu \partial_\mu \nu_{\ell \mathrm{R}} \crn
&+& \bar{L}_\ell  i \gamu \:(\partial_\mu+i\frac{g'}{2}B_\mu-ig
\frac{\tau_a}{2}W_{\mu a})\: L_\ell
\ea
using standard notation. As usual the chiral fields
\ba
\ell_{\mathrm{R}}=\Pi_+ \ell\;,\;\;\nu_{\ell \mathrm{R}}=\Pi_+
\nu_\ell\;,\;\;
L_\ell=\left( \bary{c}\nu \\ \ell \eary \right)_{\mathrm{L}}=\Pi_-
\left( \bary{c}\nu \\ \ell \eary \right)
\ea
may be represented in terms of the lepton fields $\ell(x)$ and the
neutrino field $\nu_{\ell}(x)$ with the help of the chiral projectors
\be
\Pi_{\pm}\equiv \ha \:(\mathbf{1}\pm\gafi)\;.
\ee
In order that $\Pi_\pm$ are Hermitean projection operators $\gafi$
must have the properties
\be
\gafi^2=\mathbf{1}\;,\;\;\gafi^+=\gafi\;.
\ee
Furthermore, we demand $\Pi_\pm$ to be chiral projectors also for the
adjoint $\bar{\psi}=\psi^+ \gamma^0$ of a Dirac field $\psi$. This
implies
\be
\label{acnull}
\{\gamma^0, \gafi\} =0\;.
\ee
By Lorentz covariance in the 4--dimensional physical subspace
the latter condition extends to
\be
\label{acphys}
\{\gamu , \gafi\} =0 \;\;{\mathrm{for}}\;\;\;\mu=0,1,2,3\;.
\ee
It is easy to verify that ${\cal L}_\ell$ is invariant under local
$SU(2)_{\rm L} \otimes U(1)_{\rm Y}$ gauge transformations,
irrespective of $\ac{\mu} \neq 0$. Since the chiral fields have the
simple transformation properties
\ba
L_\ell &\ra& \exp \{-i/2 \:(g' \beta-g \tau_a \omega_a)\:\Pi_-\} L_\ell=
\exp \{-i/2 \:(g' \beta-g \tau_a \omega_a)\} L_\ell \crn
\ell_{\rm R} &\ra& \exp \{-i g' \beta \Pi_+\} \ell_{\rm R}=
\exp \{-i g' \beta \} \ell_{\rm R} \crn
\nu_{\rm R} &\ra& \nu_{\rm R}\;,
\ea
the invariance of ${\cal L}_\ell$ follows immediately from the
properties of $\Pi_\pm$ alone.

We notice that in utilizing chiral fields there seems to be no
conflict with the non--anti-commutativity of $\gafi$ and the formal
validity of the ST--identities.

Usually, one prefers to write Feynman rules in terms of the Dirac
fields $\ell$ and $\nu_ \ell$. The standard Feynman rules are obtained
using the relations
\be
\bar{\psi}\Pi_\mp \gamu \Pi_\pm \psi=\bar{\psi} \gamu \Pi_\pm \psi\;,
\label{naive}
\ee
which are valid only, provided $\ac{\mu}=0$.

If $\ac{\mu} \neq 0$ in $D\neq 4$ dimensional space--time, the above
relations no longer hold and hence the standard Feynman rules manifestly
violate gauge invariance.  The correct relations, replacing (\ref{naive}),
read
\be
\bar{\psi}\Pi_\mp \gamu \Pi_\pm \psi=
\ha \bar{\psi}\:\left(\Gamma^\mu \pm \Gamma^\mu_5 \right)\: \psi\;,
\label{impro}
\ee
with
\be
\Gamma^\mu \equiv \gamu - \ha \ac{\mu} \gafi =
\ha \left(\gamu-\gafi \gamu \gafi \right)
\label{Gamu}
\ee
and
\be
\Gamma^\mu_5 = \ha [\gamu, \gafi] = \Gamma^\mu \gafi\;.
\label{Gafi}
\ee
We notice that by definition all $\Gamma$'s are anticommuting with $\gafi$
\be
\{\Gamu,\gafi\} \equiv 0\;.
\ee
According to (\ref{impro}) the proper expressions for the vector current and
for the axial--vector current read
\be
V^\mu (x)=\bar{\psi} \Gamu \psi=\bar{\psi}\gamu \psi
-\ha \bar{\psi}\ac{\mu}\gafi \psi
\label{newvec}
\ee
and
\be
A^\mu (x)=\bar{\psi} \Gamu \gafi \psi=\bar{\psi}\gamu \gafi \psi
-\ha \bar{\psi}\ac{\mu} \psi\;,
\label{newaxi}
\ee
respectively. It might be worthwhile to point out that the standard
form of the axial current $\bar{\psi}\gamu \gafi \psi $ is not
Hermitean when $\ac{\mu}\neq 0 $. The above consideration also shows
how  anomalies may come about in the vector current when $\gafi \gamu
\gafi \neq -\gamu$.

The fermion kinetic term changes to
\be
\bar{\psi}i \Gamu \partial_\mu \psi=\bar{\psi}i \gamu \partial_\mu \psi
-\ha \bar{\psi}i\ac{\mu}\gafi \partial_\mu \psi\;.
\ee
Correspondingly, the free massless fermion fields must satisfy the field
equation
\be
\left(\gamu - \ha \ac{\mu} \gafi\right) \partial_\mu \psi = 0\;.
\ee
This formally implies that the conserved canonical Noether currents
are precisely the ones given above.

By the field equation the fermion spinors satisfy
\ba
\left(\sla{k} - \ha \ac{k} \gafi -m\right)\:u(k,s) &=& 0\crn
\left(\sla{k} - \ha \ac{k} \gafi +m\right)\:v(k,s) &=& 0
\ea
and the free fermion propagator reads ($\ac{k}\equiv k_\mu \ac{\mu}$)
\be
S_F(k)=\frac{1}{\sla{k} - \ha \ac{k} \gafi -m+i0}=
\frac{\sla{k} - \ha \ac{k} \gafi +m}{K^2-m^2+i0}
\label{newprop}
\ee
with
\be
K^2\equiv k^2-\frac14 \ac{k}\ac{k}\;.
\label{sp}
\ee
Formally, we have obtained chiral and gauge invariant Feynman rules
for non-anticommuting $\gafi $. Eqs. (\ref{newvec}), (\ref{newaxi})
and (\ref{newprop}) replace the standard expressions valid for
$\ac{\mu}=0$.  

\section{Non-existence of a chirally invariant DR}

The gauge invariant Feynman rules presented in the preceding section
do not permit a regularization by continuation in the dimension $D$
when $\ac{\mu}$ is chosen compatible with the trace condition
(\ref{trcond}). This can be proven as follows. First we consider the
Dirac algebra extended to $D=2n$ ($n\geq 2$, integer). In this case
$2^n$--dimensional representations of the $\gamma$--algebra are well
known~\cite{Akye}. A basis for the algebra is given by the set of
matrices $\mathbf{1},\;\gafi$ and the antisymmetrized products
$\gamma^{[\mu_1 \ldots \mu_p]}$ associated with $p$--dimensional
subspaces of $M_D $. We will split the $SO(1,D-1) $ vectors (tensors)
into 4--dimensional vectors
$p^\mu_{\parallel}=\hat{p}^\mu\;\;(\mu=0,1,2,3) $, in the physical
subspace $M_4$, and their orthogonal
complements $p^\mu_{\perp}=\bar{p}^\mu
\;\;(\mu=4,\ldots,D-1)$. If we impose the trace condition 
(\ref{trcond}) in the physical subspace (see Eq.~(\ref{acphys}) above)
we obtain the 't~Hooft--Veltman algebra~\cite{tHVD}:
\be
\ac{\mu}=\left\{ \bary{cl} 0 & ;\; \mu=0,1,2,3 \\
                           2\bar{\gamma}^\mu \gafi & ;\; \mu=4,\ldots,D-1
                 \eary \right.
\label{tHValg}
\ee
with $\gafi = \frac{i}{4!} \veps_{\mu \nu \rho \sigma}
\hat{\gamma}^\mu \hat{\gamma}^\nu \hat{\gamma}^\rho \hat{\gamma}^\sigma$.
Here, it is important to notice that $\ac{\mu} $ is a matrix of rank
$\bar{\eps}\equiv D-4 $. The matrix--elements themselves are of order
$O(1)$. As a consequence higher products of AC-terms are
\underline{not} of higher order in $\bar{\eps} $ for $D \ra 4 $.
This is the reason why the extra terms needed to restore the \WTis
cannot be considered as perturbations. They affect the free part of
the Lagrangian! and hence the form of the fermion propagators, as
shown above. The symmetry at the end can only be there if the free and
the interacting parts of the Lagrangian match appropriately.

We are now ready to reconsider the fermion propagator
(\ref{newprop}). Using (\ref{tHValg}), we get for the scalar product
(\ref{sp})
\be
K^2=k^2-\bar{k}^2=\hat{k}^2
\ee
and thus
\be
S_F(k)=\frac{\sla{\hat{k}}+m}{\hat{k}^2-m^2+i0}
\ee
takes its 4--dimensional form, independent of $D$! It is then
impossible to regularize fermion-loop integrals by continuation in
$D$.  The crucial point is that the consistency with the trace
condition requires that in (\ref{sp}) the extra term proportional to
$\ac{k}^2$ like AC is a matrix of rank $\bar{\eps}\equiv D-4 $ and not
a correction of order $O(\eps^2)$ in the $\eps$--expansion!

The problem may be reconsidered in terms of the $\Gamma$--algebra defined by
(\ref{Gamu}), which may be associated to \underline{any} $\gamma$--algebra:
\ba
\{ \Gamu , \Gamma^\nu \} = 2 g^{\mu \nu} \cdot \mathbf{1}
-\frac14 \{\ac{\mu},\ac{\nu}\}\;,\;\;
\{\Gamu , \gafi \} = 0 \;.
\ea
For any $\Gamma$--algebra in order to be closed, we must require
\ba
\{\Gamu , \Gamma^\nu \} &=& 2 G^{\mu \nu} \cdot \mathbf{1}
\ea
for some symmetric $D\times D$--matrix $G$, which satisfies
\be
g^\mu_{~~\nu} G^{\nu \rho}=G^{\mu \rho}\;.
\ee
The trace condition (\ref{trgene}) must hold with the replacements
\be
\left(\gamu,\; g^{\mu \nu},\; \ac{\mu}\neq 0 \right) \ra
\left(\Gamu,\; G^{\mu \nu},\; \ac{\mu} = 0 \right)
\ee
which implies
\be
g_{\mu \nu} G^{\mu \nu} =G^\mu_{~~\mu}=4\;.
\ee
Assuming $G$ to have block--diagonal form
\be
G=\left(\begin{tabular}{c|c} $\hat{g}$ & 0 \\
                              \hline
                                   0 & $\bar{g}$ \end{tabular}\right)
\ee
the condition (\ref{trcond}) can be satisfied with a singular metric
$G$ only:
\be
\bar{g}=0\;,\;\; G=\hat{g}
\ee
where $\hat{g}$ must be the Minkowski metric. Thus, starting from the
't~Hooft--Veltman scheme, we are lead to a dimensional reduction (DRED)
scheme~\cite{Siegel79} by adding just some terms in the Feynman rules
which vanish in $D=4$.

As a result, the $\Gamma $--form of the 't~Hooft--Veltman algebra is
identical to the 4--dimensional Dirac algebra. In other words, using
the 't~Hooft--Veltman algebra (in its $D$--dimensional form) together
with the chiral fields, which are adapted to the gauge symmetry,
``non-regularization'' of fermion--loops is implied. Again, a
regularization can only be obtained by giving up either the trace
condition (\ref{trcond}) or gauge invariance.

This last statement, of course, is not terribly new. What we have
shown is that the Dirac algebra assuming anticommuting $\gafi $ on the
one hand and the 't~Hooft--Veltman algebra on the other hand are not
really different, since the latter can always be rewritten in the
anticommuting $\Gamma $--form by means of the relations (\ref{Gamu})
and (\ref{Gafi}).  In any case, for theories involving $\gafi $,
``dimensional regularization'' compatible with (\ref{trgene}), does
not provide well--defined integrals for loops involving fermion
lines. This has been noticed by 't~Hooft and Veltman in their original
paper~\cite{tHVD} where they state: ``the usual ambiguity of choice of
integration variables is replaced in our formalism by the ambiguity of
location of $\gafi $ in the trace''. Statements to the contrary,
frequently found in the literature, are misleading. Usually, extra
``prescriptions'' about where to put the $\gafi$ in a particular
calculation are proposed. These prescriptions, however, do not resolve
the problem of mathematical inconsistencies, i.e., they still require
an explicit check and the restoration of the \WTisp 

The use of chiral fields provides an unambiguous rule for the proper
location of the $\gafi $--matrices before generalization to $D\neq
4$. Unfortunately, this has lead to the ``non-regularization'' by
dimensional continuation when the $D\neq 4$ trace condition
(\ref{trcond}) is imposed, which in turn essentially implies the 't
Hooft--Veltman scheme.

If we violate gauge invariance by the naive application of the 't
Hooft--Veltman prescription, we have to restore the symmetry by
imposing the relevant Ward--Takahashi identities and fixing
appropriate counter terms. But this precisely amounts to including the
extra $\ac{\mu}$ terms given in Eqs.~(\ref{newvec}) and
(\ref{newaxi}). Which in turn is nothing but another way of utilizing
the naive anticommuting $\gafi$.

\section{Conclusion for the practitioner}
\label{Sec:conc}

According to our considerations above we are left with two possible
strategies:
\subsection*{i) $\ac{\mu}\neq 0$: the chirally improved 't~Hooft--Veltman
scheme} If we insist on the trace condition (\ref{trcond}) the gauge
invariance must be manifestly broken in order to obtain the ``pseudo
regularization'' by dimensional continuation. Again we start at the
level of the chiral fields but must avoid the non--regularization by
treating the AC--terms in the free part of the Lagrangian as
interaction terms, i.e., we use the standard $D$--dimensional Fermi
propagator
\be
\label{sfddim}
S_F(k)=\frac{\sla{k}+m}{k^2-m^2+i0}
\ee
together with the chiral currents (\ref{newvec}, \ref{newaxi})
as our ``chiral Feynman rules''.  Since $\ac{\mu}\neq 0$, the
choice of the Fermi propagator (\ref{sfddim}) amounts to adding the
symmetry breaking term
\be
\Delta {\cal L}_{\mathrm{SB}}=\ha
\bar{\psi} i \ac{\mu} \gafi \partial_\mu \psi =
\bar{\psi}i \bar{\gamma}^\mu \partial_\mu \psi
\ee
to the Lagrangian. Besides the fact that this operator has no
4--dimensional representation, it is not a higher order term for
$D\neq 4 $ as it would be necessary for treating it as a counter-term
(perturbation). Expanding $\Delta {\cal L}_{\mathrm{SB}} $
perturbatively amounts to the assumption that $\ac{\mu}=O(\eps) $ in
the sense of matrix elements, which conflicts with (\ref{trcond}). As
we have mentioned earlier, (\ref{trcond}) requires $\ac{\mu} $ to be a
matrix of \underline{rank} $\bar{\eps}=D-4 $ with matrix elements of
order $O(1)$. A mathematically satisfactory way out of the dilemma
within the framework of DR is not possible as a result of the
existence of the ABJ--anomaly.

Our considerations show that ``quasi gauge invariant'' Feynman rules
may be obtained for non-anticommuting $\gafi $ provided $\ac{\mu} $ is
treated as a perturbation i.e. $\ac{\mu}=O(\eps) $. Examples are
briefly considered in the Appendix. Results turn out to be
AC--independent in this case. \underline{AC--invariance} may be used
as a helpful tool for checking the gauge invariance of fermionic loop
contributions to amplitudes. Usually such checks are possible only by
explicit consideration of WT- and/or ST-identities. We stress, once
again, that any approach which treats the AC--term as a perturbation
conflicts with the trace condition (\ref{trcond}) at some
point. Ignoring this point leads to ``standard'' confusions,
frequently appearing in the literature. While working with the 't
Hooft--Veltman prescription in the standard form requires the
subsequent check of the \WTisc after utilizing the chiral version of
the Feynman rules we may restrict ourselves to check the hard anomaly
diagrams.

Since amplitudes exhibiting spurious anomalies only may be chiralized
either by our chirally improved Feynman rules or by imposing the \WTis
which makes them AC--invariant we obviously may directly choose the
scheme $\ac{\mu}=0$, which is our second and preferred option:

\subsection*{ii) $\ac{\mu}=0$: the quasi self-chiral scheme}
From a practical point of view an acceptable computational
scheme should avoid spurious anomalies in the first place. This is
possible only if the trace condition (\ref{trcond}) is given up. Gauge
invariance can be preserved then by using an anticommuting $\gafi
$. This has been noticed in~\cite{BGL} (see
also~\cite{GD79,ET82,AZ87,DK94}). 

We observe that taking chiral fields seriously on a formal level, the
only consistent way to avoid the above non-regularization is the
simple one: use anticommuting $\gafi$ from the very beginning, i.e.,
choose the NDR algebra (\ref{NDR}). Since $\Gamu \equiv \gamu$ in this
case we do not get the non-regularization of the fermion
propagators. The ABJ--anomaly must be
considered separately as we are going to discuss now~\footnote{The
terminology introduced in~\cite{BM77,GB80} which calls a scheme
``consistent'' if it respects the trace condition (\ref{trcond}) and
``inconsistent'' otherwise is definitely misleading by the
considerations presented in this paper. Since we cannot satisfy the
\WTis and the trace condition simultaneously we have the choice which
one we want to consider more fundamental. Something has to be restored
at the end by hand in any case.  To put into place the model
independent ABJ--anomalies, is by far simpler, than restoring the
chiral symmetry which is broken by non--NDR schemes.}.
 
\begin{figure}[htb]
\centerline{
% diagram 1
\begin{picture}(120,60)(-60,0)
{\SetScale{3.0}
{\SetColor{Red} 
\Photon(00,30)(15,30){1}{4}
\Photon(28,37.5)(43,37.5){1}{4}
\Photon(28,22.5)(43,22.5){1}{4}}
\put(00,28){\makebox(0,0)[t]{$-(p_1+p_2),\lambda$}}
\put(43,46){\makebox(0,0)[t]{$p_1,\mu$}}
\put(43,20){\makebox(0,0)[t]{$p_2,\nu$}}
{\SetColor{Green}
\ArrowLine(15,30)(28,37.5)
\ArrowLine(28,37.5)(28,22.5)
\ArrowLine(28,22.5)(15,30)}
\Vertex(15,30){1}
\Vertex(28,37.5){1}
\Vertex(28,22.5){1}
\put(58,30){\makebox(0,0)[]{$+$}}}
\end{picture}
\hfill
% diagrams 2
\begin{picture}(120,60)(-20,0)
{\SetScale{3.0}
{\SetColor{Red} 
\Photon(00,30)(15,30){1}{4}
\Photon(28,37.5)(43,22.5){1}{5.7}
\Photon(28,22.5)(43,37.5){1}{5.7}}
\put(00,28){\makebox(0,0)[t]{$-(p_1+p_2),\lambda$}}
\put(43,46){\makebox(0,0)[t]{$p_1,\mu$}}
\put(43,20){\makebox(0,0)[t]{$p_2,\nu$}}
{\SetColor{Green}
\ArrowLine(15,30)(28,37.5)
\ArrowLine(28,37.5)(28,22.5)
\ArrowLine(28,22.5)(15,30)}
\Vertex(15,30){1}
\Vertex(28,37.5){1}
\Vertex(28,22.5){1}}
\end{picture}
\hfill
}
\medskip
\caption[]{The VVA triangle diagrams.}
\label{fig:VVA}
\end{figure}
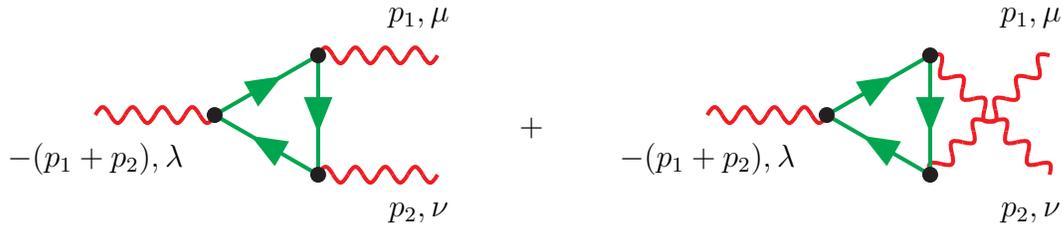

In the gauge invariant approach, closed fermion loops exhibiting
$\gafi $ odd traces and hard anomalies, cannot be obtained
by dimensional continuation, merely, $\gafi $ odd traces are to be
considered as intrinsically 4-dimensional quantities. Since charge
conjugation properties and the related Bose symmetry are not
automatically satisfied one has to account left- and right-circulation
of the fermions in closed loops separately. In any case Adler's
approach~\cite{Adler70} can be utilized to resolve the remaining
ambiguities. For this purpose, let us briefly consider the
ABJ--anomaly~\cite{ABJ} exhibited by the current
correlator $<0|T\{V^\mu(x_1) V^\nu(x_2) A^\lambda(y)\}|0>$ of two
vector currents and an axial--vector current. The one--loop diagrams
are shown in Fig.~\ref{fig:VVA}.

In $D=4$, working as usual in momentum space, we may perform a
covariant decomposition of the third rank pseudotensor which depends
on the two independent momenta $p_1$ and $p_2$:
\ba
A^{\mu\nu\lambda}(p_1,p_2)&=&
\veps^{\mu\nu\lambda \al}\:\left(p_{1\al}\:A_1 +p_{2\al}\:A_2\right) \crn
&+&\veps^{\mu \lambda \alpha \beta}\:p_{1\alpha}p_{2\beta}\:
\left(p^{\nu}_1\:A_3 + p^{\nu}_2\:A_4 \right) \crn
&+&\veps^{\nu \lambda \alpha \beta}\:p_{1\alpha}p_{2\beta}\:
\left(p^{\mu}_1\:A_5 + p^{\mu}_2\:A_6 \right) \crn
&+&\veps^{\mu \nu \alpha \beta}\:p_{1\alpha}p_{2\beta}\:
\left(p^{\lambda}_1\:A_7 + p^{\lambda}_2\:A_8 \right) 
\ea
where the amplitudes $A_i$ are Lorentz scalars. We now impose
\begin{itemize}
\item Bose symmetry (i.e. consider the sum of the two diagrams of
Fig.~\ref{fig:VVA}): 
\end{itemize}
$$A^{\mu \nu \lambda}(p_1,p_2)=A^{\nu \mu
\lambda}(p_2,p_1)$$
which implies 
\ba \begin{array}{cccccc}
A_1(p_1,p_2)&=&-A_2(p_2,p_1), &
A_3(p_1,p_2)&=&-A_6(p_2,p_1), \\ 
A_4(p_1,p_2)&=&-A_5(p_2,p_1), &
A_7(p_1,p_2)&=&+A_8(p_2,p_1).
\end{array}
\ea
\begin{itemize}
\item Vector current conservation:
\end{itemize}
$$p_{1\mu} A^{\mu\nu\lambda}=p_{2\nu} A^{\mu\nu\lambda}=0$$ 
which implies
\ba
A_1&=&-\left(p_2^2\:A_4+p_1p_2\:A_3\right) \crn 
A_2&=&-\left(p_1^2 \: A_5+p_1p_2\:A_6\right)\;.
\ea

We thus find that the amplitudes $A_1$ and $A_2$ are determined
uniquely in terms of the $A_i$, $i=3,\ldots,6$. The crucial
observation, made by Adler long time ago~\cite{Adler70}, is that the
amplitudes $A_i$, $i=3,\ldots,8$, have dimension $d_{\rm eff}=1-3=-2$
and hence are represented by convergent integrals. In contrast, $A_i$,
$i=1,2$, have dimension $d_{\rm eff}=1-1=0$ (logarithmically divergent)
and thus require regularization and renormalization. However, imposing
Bose symmetry and vector current conservation uniquely determines the
two regularization/renormalization dependent amplitudes in terms of
the other convergent and hence unambiguous ones, i.e., the result is
unique without need to refer to a specific renormalization scheme. The
divergence of the axial--vector current takes the form
$$-(p_1+p_2)_\lambda  A^{\mu\nu\lambda}=2mR^{\mu\nu}+8 \pi^2
p_{1\alpha} p_{2\beta} \veps^{\alpha\beta\mu\nu}\neq 0$$
where the first term on the r.h.s. is the normal term which vanishes
for vanishing fermion mass $m$ while the second term is the mass
independent anomaly.
Formal axial--vector current conservation in the limit of vanishing
fermion mass would require
$$A_1-A_2-(p_1^2+p_1p_2)\:A_7 -(p_2^2+p_1p_2)\:A_8=0\:? $$ with $A_1$
and $A_2$ fixed already by vector current conservation, this
expression as we know does not vanish but yields the famous
axial--vector current anomaly. All true anomalies, i.e.,quantum
effects like the triangle anomaly which cannot be removed by adding a
corresponding counter term to the Lagrangian, are well known to be
related to the triangle diagram. Besides the triangle diagram itself
they appear by tensor reduction from one--loop box and pentagon
diagrams and diagrams which contain the one--loop anomalous graphs as
subgraphs.

The Adler--Bardeen non-renormalization theorem~\cite{ABtheo} of the
one--loop anomalies implies that matters are under control provided
Bose symmetry and vector current conservation are imposed, if
necessary by hand. In DR it has been reconsidered
in~\cite{ABtheoDR,DK94}. Last but not least we must have the anomaly
cancelation, possible by virtue of the quark lepton duality, in order
to have the SM renormalizable~\cite{SManomaly}.

Summary: we have shown that different $\gafi$--schemes may be related
by adding suitable terms in the $D$--dimensional Lagrangian which
vanish at $D=4$. In any scheme we can mimic chiral fields by the
appropriate choice of the Feynman rules. We consider this to be
crucial since the physical SM derives via a Higgs mechanism from a
symmetric phase which exhibits chiral fermions only. The corresponding
``chiral completion'' (see (\ref{newvec},\ref{newaxi})) of the Feynman
rules cannot make a consistent scheme inconsistent or vice versa.
Avoidable (often called ``spurious'') anomalies are then absent.  Our
arguments strongly support the application of the NDR scheme
(\ref{NDR}), i.e., the $D$--dimensional $\gamma$--algebra together
with a strictly anticommuting $\gafi$, together with the simple
4--dimensional treatment of the hard anomalies discussed above. The
NDR is easily implemented into computer codes and is by far the most
convenient and efficient approach in calculations of radiative
corrections. Removable anomalies are avoided and hence a tedious
procedure of restoration of WT- and ST-identities is not needed.

The rules advocated here have been utilized successfully in the last
twenty years by many authors at the one-- and the two--loop level and
beyond. Most SM calculations of higher order effects adopted the NDR
scheme without encountering any inconsistencies. Of course, the NDR
scheme has been advocated by several
authors~\cite{BGL,ChFH79,GD79,ET82,AZ87,DK94} (see also~\cite{MP99})
in the past. I hope the present paper contributes to clarify part of
the ongoing controversy.

{\bf Acknowledgements } 

Part of the ideas presented here have been developed a long time ago at
the begining of the ongoing collaboration with Jochem Fleischer.  I
gratefully acknowledge in particular his involvement in the sometimes
tedious explicit checks we have performed for many SM calculations.  I
also thank Christopher Ford and Oleg Tarasov for helpful discussions
and for carefully reading the mansuscrpt.

\section*{Appendix: \\ Calculations with $\ac{\mu}\neq 0$ 
in the SM: Two examples. } We~\cite{FJ79} have verified explicitly
that all spurious anomalies disappear from fermion propagators and
fermion form factors at one-loop order for the case where we use
Feynman rules as proposed in Sec.~\ref{Sec:conc} in case $\ac{\mu}\neq
0$. As explained earlier, in order to avoid the
``non-regularization'' of fermion lines, we must treat AC as a
perturbation $\ac{\mu}=O(\eps)$ and work to linear order in AC. All
calculations have been performed in the 't~Hooft gauge with an
arbitrary gauge parameter $\xi$, which makes possible direct
analytical checks of gauge invariance. We only summarize the structure
of the results.

The irreducible self-energy $\Sigma(k)$ we obtained has the following
form
\be
\label{A:sechi}
\Sigma(k)=\left(\sla{k}-m-\ha \ac{k}\gafi \right)\:A +\ha
[\sla{k},\gafi]\:B+mC\;. 
\ee
This implies that the mass- and wave-function renormalization are
completely AC--independent:
\be
\delta m=-mc_0\;,\;\;\sqrt{Z_2}=1-\ha a_0 -\ha b_0 \gafi\,.
\ee
Here the wave-function renormalization constant is given by the matrix
\be
\sqrt{Z_2}=\sqrt{Z_R}\:\Pi_+ \:+\: \sqrt{Z_L}\: \Pi_-
\ee
where $\sqrt{Z_R}$ and $\sqrt{Z_L}$ are the independent wave--function
renormalizations of the right--handed and left--handed fields,
respectively.  Thus the renormalized self--energy reads
\be
\Sigma_r(k)=\left(\sla{k}-m-\ha \ac{k}\gafi \right)\:(A-a_0) +\ha
[\sla{k},\gafi]\:(B-b_0)+m\:(C-c_0)
\ee
with $A-a_0$, $B-b_0$ and $C-c_0$ finite, and hence
\be
\Sigma_r(k)=\left(\sla{k}-m \right)\:(A-a_0) +
\sla{k}\gafi\:(B-b_0)+m\:(C-c_0)+O(\eps)
\ee
By contrast, using standard Feynman rules, we obtain
\ba
\Sigma(k)&=&\left(\sla{k}-m \right)\:A +\ha
[\sla{k},\gafi]\:B+mC \crn
&+&\ac{k}\gafi\:D + [\sla{k},\ac{\gamma}]\:E 
+ [\sla{k},\ac{\gamma}]\:\gafi\:F
\ea
for the bare self-energy. In this case it is not possible to perform
the renormalization in the standard way without imposing the \WTis
first, which must lead to the form (\ref{A:sechi}).

Similar results can be found for form-factors. The following applies
to the $\bar{\ell}\ell\gamma$ and $\bar{\ell}\ell Z$ vertices. The
general form of the irreducible vertices reads
\be
\Pi^\mu(p_1,p_2)=\left(\gamu-m-\ha \ac{\mu}\gafi \right)\:F_1 +\ha
[\gamu,\gafi]\:F_2+p_1^\mu F_3+p_2^\mu F_4\;.
\ee
We notice that the only surviving AC--term is $\ac{\mu}\gafi$ which
appears in the canonical from (\ref{Gamu}) as in the Born term. Thus
the vertex renormalization can be performed in an AC--independent way,
i.e., the renormalized vertex is given by
\be
\Pi^\mu_r(p_1,p_2)=\left(\gamu-m-\ha \ac{\mu}\gafi \right)\:(F_1-c_1) +\ha
[\gamu,\gafi]\:(F_2-c_2)+p_1^\mu F_3+p_2^\mu F_4
\ee
with $F_1-c_1$, $F_2-c_2$, $F_3$ and $F_4$ finite. Hence, we have
\be
\Pi^\mu_r(p_1,p_2)=\gamu\:(F_1-c_1) +
\gamu\gafi\:(F_2-c_2)+p_1^\mu F_3+p_2^\mu F_4 +O(\eps)
\ee
independent of any AC--term. In contrast, by applying standard Feynman
rules, we find additional terms of the form $\ac{\mu}\gafi$,
$[\gamu,\ac{\gamma}]$ and $\{\gamu,\ac{\gamma}\}\gafi$ which cannot be
removed by renormalization, unless we impose the \WTis first.  In the
chiral scheme we obtain gauge invariant form factors directly without
imposing \WTis by hand. Calculations in this ``chiral'' scheme in fact
look very similar to the ones performed with anticommuting $\gafi$.

As a result of these findings we decided to work with an
anti-commuting $\gafi$ henceforth, first at the one--loop
level~\cite{FJ81,FJZ89}, later at the two--loop
level~\cite{FB84,FTJR92,FTJ93}. In most of these calculations we
worked in the 't~Hooft gauge with a free gauge parameter which allowed
us to check explicitly the gauge invariance of on-shell matrix
elements.

%%%%%%%%%%%%%%%%%%%%%

\bb{99}

\bibitem{SM}{
S. L. Glashow, \np 22 (1961) 579; \\
S. Weinberg, \prl {\bf 19} (1967) 1264; \\
A. Salam, in {\em Elementary Particle Theory}, ed. N. Svartholm, \\
Amquist and Wiksells, Stockholm (1969), p. 376.}

\bibitem{Sirlin99}{
A. Sirlin, in Proc. of the 
{\em 19th International Symposium on Lepton and Photon 
Interactions at High-Energies'} (LP 99),
Stanford, California, hep-ph/9912227.} 

\bibitem{tHooft71}{
G. 't Hooft, \npb {\bf 33} (1971) 173; {\bf 35} (1971) 167.}

\bibitem{tHVR}{ 
G. 't Hooft, M. Veltman, \npb {\bf 50} (1972) 318.}

\bibitem{tHVD}{
G. 't Hooft, M. Veltman, \np {\bf 44} (1972) 189.}

\bibitem{BGDR}
{C. G. Bollini, J.J. Giambiagi, A. Gonzales Dominguez, {\em Nuovo Cim.} 
{\bf 31} (1964) 551;\\
C. G. Bollini, J.J. Giambiagi, {\em Nuovo Cim.} {\bf 12A} (1972) 20.}

\bibitem{BGL}{
 W. A. Bardeen, R. Gastmans, B. Lautrup, \np {\bf 46} (1972) 319.}

\bibitem{Akye}
{D. Akyeampong, R. Delbourgo, {\em Nuovo Cim.} {\bf 17A} (1973) 47.}

\bibitem{BM77}{
P. Breitenlohner, D. Maison, {\em Commun. Math. Phys.} {\bf 52} (1977) 39.}

\bibitem{Siegel79}{
W. Siegel, \pl {\bf B84} (1979) 193; \pl {\bf B94} (1980) 37.} 

\bibitem{ChFH79}{
M. Chanowitz, M. Furman, I. Hinchliffe, \np {\bf 159} (1979) 225.}

\bibitem{NaWe79}{
O. Nachtmann, W. Wetzel, \pl {\bf B81} (1979) 211.}

\bibitem{GB80}{
G. Bonneau, \pl {\bf B94} (1980) 147; \npb {\bf 177} (1981) 461;
{\em Int. J. Mod. Phys.} {\bf A5} (1990) 3831.}

\bibitem{AT81}{
S. Aoyama, M. Tonin, \npb {\bf 179} (1981) 293.}

\bibitem{BDGKSch91}{
A. Barroso, M. A. Doncheski, H. Grotch, J. G. K\"orner, K. Schilcher,
\pl {\bf B261} (1991) 123.}

\bibitem{WT}{
J. C. Ward, \pr {\bf 78} (1950) 1824;\\
Y. Takahashi, {\em Nuovo Cim.} {\bf 6} (1957) 370.}

\bibitem{ST}{
A. Slavnov, {\em Theor. Math. Phys.} {\bf 10} (1972) 99;\\
J. C. Taylor, \np {\bf 33} (1971) 436;\\
G. 't Hooft, \np {\bf 35} (1971) 167.}

\bibitem{BRS}{C. Becchi, A. Rouet, R. Stora, {\em Comm. Math. Phys.}
{\bf 42} (1975) 127; see also:\\
B. W. Lee, Les Houches, Session XXVIII, 1975, {\em Methods in field theory},
eds. R. Balian, J. Zinn-Justin, North Holland, Amsterdam, 1976.}

\bibitem{KNSch90}{
J. G. K\"orner, N. Nasrallah, K. Schilcher, \prd {\bf 41} (1990) 888.}

\bibitem{BuWe90}{
A. J. Buras, P. H. Weisz, \npb {\bf 333} (1990) 66.}

\bibitem{La93}{
S. A. Larin, \pl {\bf B303} (1993) 113.}

\bibitem{FLOR95}{
R. Ferrari, A. Le Yaouanc, L. Oliver, J.C. Raynal, \prd {\bf 52}
(1995) 3036.}

\bibitem{ABJ}{
S. L. Adler, \pr {\bf 177} (1969) 2426;\\
J. S. Bell, R. Jackiw, Nuovo Cim. {\bf 60A} (1969) 47;\\
W. A. Bardeen, \pr {\bf 184} (1969) 1848.}

\bibitem{SManomaly}{
C. Bouchiat, J. Iliopoulos, P. Meyer, \pl {\bf 38} (1972) 519;\\
D. Gross, R. Jackiw, \prd {\bf 6} (1972) 477;\\
C. P. Korthals Altes, M. Perrottet, \pl {\bf 39} (1972) 546.}

\bibitem{GD79}{
S. Gottlieb, J.T. Donohue, \prd {\bf 20} (1979) 3378.}

\bibitem{ET82}{
D. Espriu, R. Tarrach, \zp {\bf 16} (1982) 77.}

\bibitem{Ovrut83}{
B. A. Ovrut, \npb {\bf 213} (1983) 241.}

\bibitem{AZ87}{
M. I. Adelhafiz, M. Zra\l ek, {\em Acta Phys. Pol.} {\bf B18} (1987) 21.}

\bibitem{cyclic}{
H. Nicolai, P. K. Townsend, \pl {\bf B93} (1980) 111;\\
D. Kreimer, \pl {\bf B237} (1990) 59;\\
J. G. K\"orner, D. Kreimer, K. Schilcher, \zp {\bf 54} (1992) 503.} 

\bibitem{MP99}{
M. Pernici, {\tt hep-th/9912278} and references therein.}

\bibitem{Wein00}{
S. Weinzierl, NIKHEF-99-010, {\tt hep-ph/9903380} (unpublished).}

\bibitem{Frei00}{
A. Freitas, presented at {\em Loops \& Legs 2000}, Bastei/K\"onigstein,
April 2000}

\bibitem{MLCHI}{
M. L\"uscher, \pl {\bf B428} (1998) 342; \npb {\bf 549} (1999)
295; \npb {\bf 568} (2000) 162.}

\bibitem{NNnogo}{
H. B. Nielsen, M. Ninomiya, \pl {\bf B105} (1981) 219.}

\bibitem{EKraus}{
E. Kraus, {\em Annals Phys.} {\bf 262} (1998) 155.}

\bibitem{ABtheo}{
S. L. Adler, W. A. Bardeen, \pr {\bf 182} (1969) 1517.}

\bibitem{ABtheoDR}{
D. R. T. Jones, J. P. Leveille, \npb {\bf 206} (1982) 473;\\
M. Bos, \npb {\bf 404} (1993) 215.}

\bibitem{DK94}{
D. Kreimer, UTAS-PHYS-94-01, {\tt hep-ph/9401354} (unpublished).}

\bibitem{Adler70}{
S. L. Adler, in {\em Lectures on Elementary Particles and Quantum
Field Theory} Vol. 1, p.1, eds. S. Deser, M. Grisaru, H. Pendleton,
M.I.T. Press, Cambridge, 1970.}

\bibitem{FJ79}{
J. Fleischer, F. Jegerlehner, Bielefeld University Preprint 1979, unpublished.}

\bibitem{FJ81}{
J. Fleischer, F. Jegerlehner, \prd {\bf 23} (1981) 2001; \npb {\bf
216} (1983); \npb {\bf 228} (1983) 1.}

\bibitem{FJZ89}{
J. Fleischer, F. Jegerlehner, M. Zra\l ek, \zp {\bf 42} (1989) 409.}

\bibitem{FB84}{
F. Bollermann, {\em PCAC in der QCD}, Bielefeld University Diploma Thesis,
1984, unpublished.}

\bibitem{FTJR92}{
J. Fleischer, O. V. Tarasov, F. Jegerlehner,  P. R\c{a}czka, \pl {\bf B293}
(1992) 437.}

\bibitem{FTJ93}{
J. Fleischer, O. V. Tarasov, F. Jegerlehner, \pl {\bf B319} (1993) 249;
\prd {\bf 51} (1995) 3820.}

\end{thebibliography}

\end{document}